\pdfoutput=1

\documentclass[11pt]{article}

\usepackage[final]{acl}
\usepackage{svg}

\usepackage{times}
\usepackage{latexsym}

\usepackage[T1]{fontenc}

\usepackage[utf8]{inputenc}

\usepackage{microtype}

\usepackage{inconsolata}

\usepackage{graphicx}

\usepackage{multirow}
\usepackage{tabularx}
\usepackage{amsmath}
\usepackage{amssymb}
\usepackage{hyperref}
\def \OURS {PeaPOD}

\title{Collaborative User Prompt for Personalized Generative Recommendation}

\author{Jerome Ramos \quad Bin Wu \quad Aldo Lipani\\
  University College London, London, UK \\
  \texttt{\{jerome.ramos.20, bin.wu.23, aldo.lipani\}@ucl.ac.uk}}

\begin{document}
\maketitle
\begin{abstract}

Large Language Models (LLMs) have become powerful foundations for generative recommender systems, framing recommendation tasks as text generation tasks. However, existing generative recommendation methods often rely on discrete ID-based prompts or task-specific soft prompts, which overlook the valuable collaborative signals shared among users with similar interests. To address this limitation, this paper presents a compositional framework that integrates a user's individual preferences with collective preferences from similar users to build personalized soft prompts. Specifically, an attention-based mechanism fuses embeddings from users with similar interests, creating a richer representation that captures multiple facets of user preferences. This design dynamically emphasizes shared interests while preserving individual user preferences. Experiments on three real-world datasets demonstrate the effectiveness of the proposed approach across sequential recommendation, top-n recommendation, and explanation generation tasks, underscoring the advantages of incorporating collaborative signals through an attention-based compositional strategy.


\end{abstract}

\section{Introduction}
\label{sec:introduction}

Recommender systems are essential tools that guide users through vast number of items and tailor personalized experiences across various tasks, such as sequential recommendation, top-n recommendation, and explanation generation~\cite{geSurveyTrustworthyRecommender2022}.
The emergence of large language models (LLMs) has opened new avenues for generative recommendation~\cite{deldjooGenRec2024}. Recent efforts such as P5~\cite{gengRecommendationLanguageProcessing2022} and M6-Rec~\cite{cuiM6RecGenerativePretrained2022} demonstrate the promise of LLM-based recommenders that unify diverse recommendation tasks under a single training paradigm, facilitating shared learning and effective knowledge transfer.


P5 ~\cite{gengRecommendationLanguageProcessing2022} formulated recommendation as a text-to-text sequence task and trained the model to perform generative recommendations based on discrete prompt templates.
\citet{liPromptDistillationEfficient2023} further proposed \textit{Prompt Distillation} to address the disconnect between the IDs (including user and item) used in the discrete template and their original natural language meaning. In particular, they employ a continuous prompt to capture the information of the discrete prompt in a more flexible manner. However, existing work maintains prompts that are reused for all users. This globally-shared prompt design is in conflict with the fact that users' personalized preferences vary significantly across the entire user population \citep{wu2022meta}. In the real world, users might have similar or distinct preferences behind their interactions.

One intuitive solution for personalizing generative recommender systems is to create a personalized prompt for each user based solely on their individual interactions. \citet{tan-etal-2024-democratizing} showed that fine-tuning a unique parameter-efficient fine-tuning (PEFT) module on each user’s interaction history can effectively personalize LLMs for text generation and classification. However, this method does not capture shared preferences between users with similar interests, which have been shown to improve personalization in traditional recommender systems~\cite{korenMatrixFactorizationTechniques2009, mnihProbabilisticMatrixFactorization2007, chengWideDeepLearning2016}.

Building on this insight, we propose \textbf{PErsonAlized PrOmpt Distillation (\OURS{})}, a flexible framework that combines individual and group-level signals into a unified \textit{collaborative user prompt} for recommendation. Our core contribution lies in effectively applying collaborative information from similar users—a technique proven successful in traditional recommender systems—and formulating it as soft prompts for LLM-based recommendation. Our approach begins by capturing shared user preferences through collaborative filtering, generating user embeddings that encode both personalized interests and group preferences from users with similar tastes. We then refine these embeddings using a compositional, attention-based mechanism that learns contextualized representations of user-to-user relationships. Specifically, the user embedding of target user $u$ serves as the query vector in a multi-head attention module, while the embeddings of the top-n most similar users, identified via cosine similarity, act as both key and value vectors. By dynamically weighting shared preferences among similar users, our model generates a collaborative user soft prompt that effectively captures and aligns with the target user's preferences. As a result, \OURS{} generates collaborative user prompts that can be reused across different recommendation tasks for consistent performance gains.
Across three Amazon datasets, \OURS{} achieves state-of-the-art performance on three recommendation tasks. We provide all data and code used at: \url{https://github.com/jeromeramos70/peapod}. 

Our key contributions are summarized below:

\begin{enumerate}
    \item We propose \textbf{PErsonAlized PrOmpt Distillation (PeaPOD)}, a novel compositional strategy that distills collaborative user preferences into a personalized soft prompt for LLM-based recommendation.
    \vspace{-0.5em}
    \item {We leverage the query-key-value mechanism from multi-head attention to capture diverse user-to-user relationships and construct a contextualized user representation that integrates both individual- and group-level information.}
    \vspace{-0.5em}
    \item {We conduct extensive experiments on sequential recommendation, top-n recommendation, and explanation generation tasks. Comprehensive ablation studies further validate the effectiveness and superiority of our proposed PeaPOD framework.}

\end{enumerate}

\section{Related Works}
\label{sec:related_works}
\subsection{LLMs for Recommendation}
Recent work has explored using LLMs for recommendation. The two main approaches of LLMs for recommender systems are (1) using the LLM's pre-trained knowledge to recommend items in a zero-shot or few-shot setting and (2) fine-tuning the model for a given domain.

A key advantage of LLMs is that they are pre-trained with a large amount of data from various domains. \citet{sannerLargeLanguageModels2023} crowdsourced a dataset of natural language summaries of user preferences and showed that prompting a language model with these user summaries yielded competitive performance with traditional baselines in a near cold-start setting. \citet{ramosNaturalLanguageUser2024} showed that a generative recommendation system trained solely on natural language user profiles is also competitive with traditional recommender systems in a warm-start setting, with the added benefit of transparency and scrutability.
Although prompting LLMs can lead to strong performance in recommender tasks, they rely more on content knowledge than on collaborative knowledge~\cite{heLargeLanguageModels2023, wu2024understanding}. Thus, if the item metadata were not included in the pre-trained knowledge base or if target users interact differently from those in the LLM's training data, the performance of the recommender system may suffer.

To mitigate these issues, the model can be further fine-tuned on target datasets in order to train the model based on the given user population and add/update its knowledge on the item dataset. P5~\cite{gengRecommendationLanguageProcessing2022} trained a T5-based~\cite{raffelExploringLimitsTransfer2020a} foundational recommender model to unify various recommendation tasks in a shared
framework. VIP5~\cite{gengVIP5MultimodalFoundation2023} extends P5 to a multi-modal foundational recommender system. POD~\cite{liPromptDistillationEfficient2023} builds upon the P5 model and appends a task-specific, continuous prompt to the input to improve performance. Although these models include user IDs as inputs, they still do not leverage the collaborative information from similar users. \citet{2024ZhuLLM4Rec} expands the tokenizer vocabulary to add user/item IDs as tokens and learn collaborative information, though this technique is infeasible as the number of users/items grows. In this work, we focus on developing a foundational generative recommender model that can efficiently encode individual and collaborative preferences for multiple downstream tasks.

\subsection{Structural Prompt}
One challenge in generative recommendation is effectively encoding user-item interactions for LLMs. Researchers have proposed numerous methods to bridge the gap between IDs and natural language. One option is to add the user's name and item title to the input prompt. However, such data may be unavailable or non-unique, making it difficult to personalize recommendations. Another option is to insert user and item IDs into the discrete prompt template~\cite{gengRecommendationLanguageProcessing2022}. However, LLMs are trained on natural language text rather than discrete IDs, limiting the effectiveness of this approach~\cite{liPromptDistillationEfficient2023}.

To tackle these issues, many researchers have explored how to effectively encode item IDs with LLMs. \citet{rajputRecommenderSystemsGenerative2023} proposed using RQ-VAE with item metadata to generate semantically meaningful tuples of tokens to serve as the semantic ID for each item. GPTRec uses a novel SVD tokenization method to generate quantized item embeddings based on user-item interactions~\cite{petrovGenerativeSequentialRecommendation2023}. \citet{tanIDGenRecLLMRecSysAlignment2024} proposed training a textual ID generator alongside the LLM-based recommender to convert item IDs to meaningful text descriptions. While these methods are effective, they do not address the need for encoding user preferences as input to the model.

To incorporate user preferences into the LLM-based recommender, \citet{zhang2024collmintegratingcollaborativeembeddings} propose a mapping technique to project user embeddings learned from collaborative filtering into a soft prompt. In this work, we focus on developing an attention-based compositional method to better model user-to-user relationships. Previous work in continual learning for image classification has shown that an attention-based mechanism effectively shares information between similar classes of images for better performance~\cite{smithCODAPromptCOntinualDecomposed2023}. In the context of generative recommendation, attention~\cite{Vaswani2017AttentionIA} can be used to capture both individual and collaborative preferences across diverse feature spaces. 

\section{Preliminary}
\label{sec:preliminary_work}
\subsection{Recommendation Tasks}
Following existing works~\cite{gengRecommendationLanguageProcessing2022, liPromptDistillationEfficient2023}, we focus on three important recommendation tasks: (1) Sequential Recommendation, (2) Top-n Recommendation, and (3) Explanation Generation. Let $u \in U$ denote a user in the user set and $i \in I$ denote an item in the item set across all three tasks. We aim to train a unified model to complete the following tasks:
\begin{itemize}
    \item \textit{Sequential Recommendation:} given a user $u$ and a list of historical interactions in chronological order $i_1, i_2, ... i_{n-1}$, the task is to predict the next item $i_n$ that $u$ will interact with.
    \item \textit{Top-n Recommendation:} given a pool of randomly sampled items from $I$, the model must select the top $N$ items.
    \item \textit{Explanation Generation:} given a user $u$ and item $i$, the model must generate an explanation $E_{u,i}$.
\end{itemize}


Importantly, the scope of our study focuses only on user and item IDs for recommendation and excludes additional information such as item metadata or user profiles since this information may be sensitive or unavailable.




\subsection{Prompt Distillation}
In traditional recommender systems, user and item IDs are commonly used to identify unique users and items. However, since these IDs are unique to the dataset and are not natural language text, it is necessary to perform additional fine-tuning to allow the LLM to effectively understand these IDs. The most common methodology is to train the model with \textit{discrete prompt templates}. For example, in the case of explanation generation, the discrete prompt could be ``Explain why \underline{user\_1234} enjoys \underline{item\_5678}''. When using IDs with discrete prompt templates, recent research proposes adding an additional embedding space, called \textit{whole-word embeddings}, so that the model can recognize which tokens belong to the appropriate ID~\cite{gengRecommendationLanguageProcessing2022}. That is, each token in the ID utilizes the same whole-word embedding and all other tokens are tokenized as normal.


\begin{figure*}[t]
  \centering 
  \includegraphics[width=0.8\linewidth]{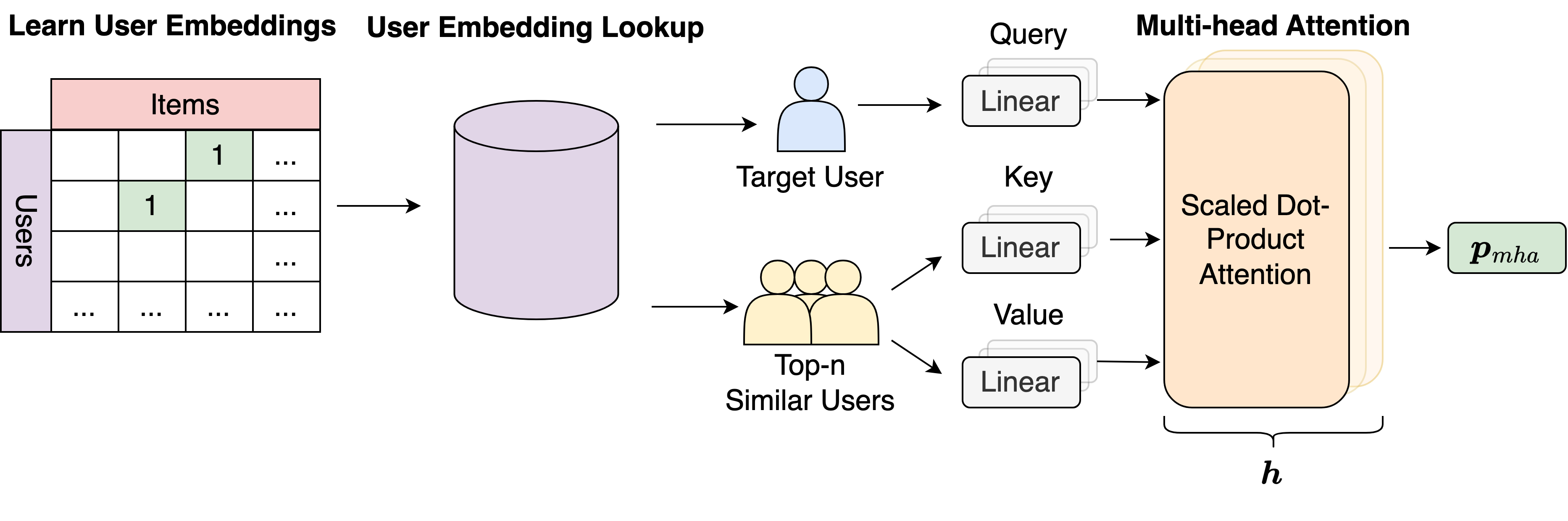}
  \caption{Full PeaPOD model architecture. We first initialize a set of user embeddings using probabilistic matrix factorization. We then apply multi-head attention, where the user embedding of target user $u$ is passed to the query vector and the embeddings of the top-n most similar users $S_u$ are passed to the key and value vectors. The final output is the collaborative user prompt $\boldsymbol{p}_{mha}$, which dynamically captures collaborative knowledge from $S_u$.}
  \label{fig:architecture}
\end{figure*}

Although discrete prompts can lead to good performance, they are inflexible and require manually crafted prompts. Thus, the performance of the model may depend strongly on how well the instructions were written. On the other hand, continuous prompts are dynamically adjusted and fine-tuned through techniques like soft prompt tuning, which optimizes prompts by learning continuous embeddings. This approach allows the model to learn regardless of the quality of the prompt template. \citet{liPromptDistillationEfficient2023} introduced \textit{Prompt Distillation}, which distills discrete prompts into continuous prompt vectors. The authors used prompt distillation to incorporate task-specific, continuous prompts for each recommendation task. These prompts are shared globally across all users to improve the model's performance. Since this method adds a shared continuous prompt per recommendation task, we will refer to these continuous prompts as \textit{task-specific prompts}. Although task-specific prompts were shown to improve performance in our recommendation tasks, they cannot express personalization because each user utilizes the same set of prompts. Thus, we develop a new method to distill each user's preferences into a \textit{personalized} soft prompt that caters to each user's individual needs, while incorporating relevant preferences from similar users.


\section{Collaborative User Prompt}
\label{sec:methodology}

The goal of the \textit{collaborative user prompt} is to distill a user's preferences into a soft prompt to represent the user's personal interests while also leveraging collaborative knowledge from users with similar preferences. To accomplish this, we implement multi-head attention~\cite{Vaswani2017AttentionIA} to create a contextualized user representation that combines individual and group-level information. Attention dynamically weights the influence of each similar user, producing a richer and more robust representation. Furthermore, multiple heads allows the model to capture different aspects of similarity, resulting in a more comprehensive and diverse understanding of the target user. By applying multi-head attention, we generate a collaborative user prompt that can be effectively used and shared by the LLM across multiple recommender tasks.

\subsection{{User Embedding Representation}}
To generate the collaborative user prompt, we must first select an effective input for the multi-head attention module. In our case, we utilize user embeddings generated from probabilistic matrix factorization~\cite{mnihProbabilisticMatrixFactorization2007} for each user. Specifically, given a user-item feedback matrix $\boldsymbol{F} \in \mathbb{R}^{|\boldsymbol{U}| \times |\boldsymbol{I}|}$, the model learns a user matrix $\boldsymbol{U} \in \mathbb{R}^{|\boldsymbol{U}| \times d_u}$ and item matrix $\boldsymbol{I} \in \mathbb{R}^{|\boldsymbol{I}| \times d_u}$, where $d_u$ is the selected dimension size of the user embedding and $|\boldsymbol{U}|$ and $|\boldsymbol{I}|$ are the number of users and items, respectively. $\boldsymbol{u} \in \boldsymbol{U}$ represents a user embedding that encodes a user's individual preferences. By construction, users with similar preferences are positioned closer to each other in this embedding space (e.g. higher cosine similarity). These pre-trained user embeddings serve as a strong initialization for personalization because they already capture collaborative signals.

\subsection{Attending User Embeddings}
\label{subsec:attention_user_embeddings}
In our framework, we utilize attention to generate a collaborative user prompt $\boldsymbol{p}$ by leveraging the embeddings of user $u$ (e.g. $\boldsymbol{u}$) and $\boldsymbol{S}_u \in \mathbb{R}^{n \times d_u}$, which consists of the top-n most similar user embeddings to $\boldsymbol{u}$ according to cosine similarity. We attend to these $n$ neighboring user embeddings to capture their collective preferences, conditioned on $u$. We first project $\boldsymbol{u}$ and each neighbor's embedding $\boldsymbol{s}_i \in \boldsymbol{S}_u$ into the query $\boldsymbol{Q}$, key $\boldsymbol{K}$, and value $\boldsymbol{V}$ subspaces.
\begin{align}
\boldsymbol{Q} &= \boldsymbol{W}_Q \boldsymbol{u} + \boldsymbol{b}_Q \in \mathbb{R}^{d_m} \\
\boldsymbol{K}_i &= \boldsymbol{W}_K \boldsymbol{s}_i + \boldsymbol{b}_K \in \mathbb{R}^{d_m} \\
\boldsymbol{V}_i &= \boldsymbol{W}_V \boldsymbol{s}_i + \boldsymbol{b}_V \in \mathbb{R}^{d_m}
\end{align}
where $d_m$ is the dimensionality of the base LLM. We aggregate the key and value vectors to form the key and value spaces $\boldsymbol{K} \in \mathbb{R}^{n \times d_m}$ and $\boldsymbol{V} \in \mathbb{R}^{n \times d_m}$, respectively.

Let $\boldsymbol{Q}$ represent the target user $u$ on which we condition group-level information. $\boldsymbol{K}$ encodes $\boldsymbol{S}_u$ in the key space, enabling the model to learn how relevant each similar user is to $u$. $\boldsymbol{V}$ carries the content from $\boldsymbol{S}_u$ to be aggregated by attention. We use scaled dot-product attention to compute an attention vector and apply it to $\boldsymbol{V}$:
\begin{equation}
 \text{Attn}(\boldsymbol{Q},\boldsymbol{K},\boldsymbol{V}) = \text{softmax}\left(\frac{\boldsymbol{Q}\boldsymbol{K}^\top}{\sqrt{d_m}}\right)\boldsymbol{V}
\end{equation}
This results in an attention-weighted summary vector, which we define as $\boldsymbol{z}$. This allows us to dynamically aggregate the most important information from similar users with respect to $u$, further enhancing the collaborative information learned from the initialized user embeddings. Finally, we apply a linear layer to $\boldsymbol{z}$ to obtain the collaborative user prompt $\boldsymbol{p}$.

\begin{align}
    \boldsymbol{p} &= \boldsymbol{W}_{l} \boldsymbol{z} + \boldsymbol{b}_{l} \in \mathbb{R}^{d_p}
\end{align}
where $d_p$ is the selected prompt length for $\boldsymbol{p}$. By attending to the top-n most similar users, we dynamically incorporate relevant collaborative signals into $\boldsymbol{p}$ based on $u$. This compositional strategy combines group-level information, conditioned on $u$'s preferences, to distill collaborative preferences into a personalized soft prompt.

The collaborative user prompt described in Section~\ref{subsec:attention_user_embeddings} can be extended from a single-head implementation to a multi-head attention~\cite{Vaswani2017AttentionIA} to allow each head to focus on different aspects of the user embeddings. We show the full architecture to obtain the multi-head prompt $p_{mha}$ in Figure~\ref{fig:architecture}.


\subsection{Training and Inference}
Using the task-alternated training strategy proposed by~\citet{liPromptDistillationEfficient2023}, we alternate the task being trained for each batch. This allows us to save time during training because the samples in the batch have the same data format and will therefore be of similar length to one another. Importantly, task-alternated training did not show any degradation in generative recommendation performance versus traditional random batching strategies~\cite{liPromptDistillationEfficient2023}. We combine collaborative user prompts, task-specific prompts, and discrete prompts as input during training. We also add whole-word embeddings to help the model identify which tokens belong to a particular ID~\cite{gengRecommendationLanguageProcessing2022}.

As all the generative recommendation tasks are posed as sequence-to-sequence tasks, we use the Negative Log-Likelihood (NLL) loss to optimize the model. In addition, the model must output a sequence of text that forms an item ID (or natural language text in the case of explanation generation). For these tasks, we choose beam search, which is commonly used for its effectiveness in sequence-to-sequence generation.

For sequential recommendation and top-n recommendation, the $b$ candidate sequences form the recommendation list, where $b$ is the beam width. For explanation generation, we select the sequence with the largest log-likelihood from the candidates.

\section{Experimental Setup}
\label{sec:experimental_setup}
The following research questions guide the remainder of the paper:
\begin{itemize}
    \item \textbf{RQ1:} How does our proposed \OURS{} model compare with existing state-of-the-art models on different tasks?
    \vspace{-0.5em}
    \item \textbf{RQ2:} What are the effects of adding a collaborative user prompt as input?
    \vspace{-0.5em}
    \item \textbf{RQ3:} How useful is multi-head attention vs. alternative implementations such as single-head attention or multi-layer perceptron?
    \vspace{-0.5em}
    \item \textbf{RQ4:} How does model size impact performance?
\end{itemize}
\subsection{Datasets}
\label{subsec:datasets}

For our experiments, we use the Sports, Beauty, and Toys \& Games Dataset from Amazon~\cite{ni-etal-2019-justifying}. We reuse the preprocessing steps and the 8:1:1 train/validation/test splits created by~\citet{gengRecommendationLanguageProcessing2022}. Each sample consists of a user ID, an item ID, a text review, a rating, and a time stamp. We provide the dataset statistics in Table~\ref{tab:datasets}.

\begin{table}[t]
\centering
\begin{tabular}{lrrrr}
\hline
\textbf{Dataset} & \textbf{Sports} & \textbf{Beauty} & \textbf{Toys} \\
\hline
\#Users & 35,598 & 22,363 & 19,412 \\
\#Items & 18,357 & 12,101 & 11,924 \\
\#Reviews & 296,337 & 198,502 & 167,597 \\
Sparsity (\%) & 0.0453 & 0.0734 & 0.0724 \\
\hline
\end{tabular}
\caption{Statistics of the datasets.}
\label{tab:datasets}
\end{table}
\begin{table*}[t]
\centering
\caption{Performance comparison on sequential recommendation, evaluated using Hit Rate (HR@k) and NDCG (N@k).}
\label{tab:sequential_performance_comparison}
\resizebox{\textwidth}{!}{
\begin{tabular}{lcccc|cccc|cccc}
\hline
\multirow{2}{*}{Methods} & \multicolumn{4}{c}{Sports} & \multicolumn{4}{c}{Beauty} & \multicolumn{4}{c}{Toys} \\
\cline{2-5} \cline{6-9} \cline{10-13}
 & HR@5 & N@5 & HR@10 & N@10 & HR@5 & N@5 & HR@10 & N@10 & HR@5 & N@5 & HR@10 & N@10 \\
\hline
GRU4Rec & 0.0129 & 0.0086 & 0.0204 & 0.0110 & 0.0137 & 0.0099 & 0.0283 & 0.0137 & 0.0097 & 0.0059 & 0.0176 & 0.0084 \\
BERT4Rec & 0.0115 & 0.0075 & 0.0191 & 0.0099 & 0.0203 & 0.0124 & 0.0347 & 0.0170 & 0.0116 & 0.0071 & 0.0203 & 0.0099 \\
FDSA & 0.0182 & 0.0122 & 0.0288 & 0.0156 & 0.0267 & 0.0163 & 0.0407 & 0.0208 & 0.0228 & 0.0140 & 0.0381 & 0.0189 \\
SASRec & 0.0233 & 0.0154 & 0.0350 & 0.0192 & 0.0387 & 0.0249 & 0.0605 & 0.0318 & 0.0463 & 0.0306 & 0.0675 & 0.0374 \\
S$^3$-Rec & 0.0251 & 0.0161 & 0.0364 & 0.0204 & 0.0387 & 0.0247 & 0.0607 & 0.0327 & 0.0443 & 0.0294 & 0.0640 & 0.0376 \\
P5 & 0.0272 & 0.0169 & 0.0361 & 0.0198 & 0.0530 & 0.0370 & 0.0659 & 0.0421 & 0.0460 & 0.0567 & 0.0709 & 0.0587 \\
VIP5 & 0.0412 & 0.0345 & 0.0475 & 0.0365 & 0.0556 & 0.0427 & 0.0677 & 0.0467 &  0.0662 & 0.0577 & 0.0749 & 0.0604 \\
IDGenRec & 0.0429 & 0.0326 & 0.0574 & 0.0372 & 
\underline{0.0618} & \textbf{0.0486} & \underline{0.0814} & \textbf{0.0541} & 0.0655 & 0.0481 & \underline{0.0870} & 0.0551 \\
POD & 0.0496 & 0.0396 & 0.0576 & 0.0419 & 0.0537 & 0.0395 & 0.0688 & 0.0443 & 0.0691 & \underline{0.0599} & 0.0742 & \underline{0.0610} \\
RDRec & \underline{0.0505} & \underline{0.0408} & \underline{0.0596} & \underline{0.0433} & 0.0601 & 0.0461 & 0.0743 & 0.0504 & \underline{0.0723} & 0.0593 & 0.0802 & 0.0605 \\
\hline
PeaPOD
& \textbf{0.0554*} & \textbf{0.0440*} & \textbf{0.0665*} & \textbf{0.0473*} &
\textbf{0.0649} & \underline{0.0469} & \textbf{0.0843} & \underline{0.0536} & 
\textbf{0.0783*} & \textbf{0.0641*} & \textbf{0.0917*} & \textbf{0.0677*} \\
\hline
\end{tabular}
}
\end{table*}
\begin{table*}[t]
\centering
\caption{Performance comparison on top-n recommendation, evaluated using Hit Rate (HR@k) and NDCG (N@k).}
\label{tab:topn_performance_comparison}
\resizebox{\textwidth}{!}{
\begin{tabular}{lccccc|ccccc|ccccc}
\hline
\multirow{2}{*}{Methods} & \multicolumn{5}{c}{Sports} & \multicolumn{5}{c}{Beauty} & \multicolumn{5}{c}{Toys} \\
\cline{2-16}
 & HR@1 & HR@5 & N@5 & HR@10 & N@10 & HR@1 & HR@5 & N@5 & HR@10 & N@10 & HR@1 & HR@5 & N@5 & HR@10 & N@10 \\
\hline
 MF & 0.0314 & 0.1404 & 0.0848 & 0.2563 & 0.1220 & 0.0311 & 0.1426 & 0.0857 & 0.2573 & 0.1224 & 0.0233 & 0.1066 & 0.0641 & 0.2003 & 0.0940 \\
MLP & 0.0351 & 0.1520 & 0.0927 & 0.2671 & 0.1296 & 0.0317 & 0.1392 & 0.0848 & 0.2542 & 0.1215 & 0.0252 & 0.1142 & 0.0688 & 0.2077 & 0.0988 \\
P5 & 0.0567 & 0.1514 & 0.1049 & 0.2196 & 0.1269 & 0.0571 & 0.1566 & 0.1078 & 0.2317 & 0.1318 & 0.0451 & 0.1322 & 0.0889 & 0.2023 & 0.1114 \\
VIP5 & 0.0699 & 0.1882 & 0.1304 & 0.2717 & 0.1572 & 0.0615 & 0.1655 & 0.1147 & 0.2407 & 0.1388 & 0.0433 & 0.1301 & 0.0875 & 0.2037 & 0.1110\\
POD & 0.0895 & 0.2086 & 0.1506 & 0.2873 & 0.1756 & 0.0829 & 0.1926 & 0.1391 & 0.2670 & 0.1629 & 0.0567 & 0.1433 & 0.1009 & 0.2082 & 0.1215 \\
RDRec & \underline{0.1285} & \underline{0.2747} & \underline{0.2033} & \textbf{0.3683} & \underline{0.2326} & \underline{0.1203} & \underline{0.2572} & \underline{0.1902} & \underline{0.3380} & \underline{0.2160} & \underline{0.0660} & \underline{0.1655} & \underline{0.1171} & \underline{0.2375} & \underline{0.1398} \\
\hline
PeaPOD & 
\textbf{0.1315} & \textbf{0.2784} & \textbf{0.2075} & \underline{0.3645} & \textbf{0.2349} 
& \textbf{0.1217} & \textbf{0.2580} & \textbf{0.1915} & \textbf{0.3417} & \textbf{0.2189} 
& \textbf{0.0933*} & \textbf{0.2033*} & \textbf{0.1498*} & \textbf{0.2793*} & \textbf{0.1738*} \\

\hline
\end{tabular}
}
\end{table*}

\subsection{Baselines}
For our baselines, we selected traditional models and multi-task LLM-based recommenders for each task as well as LLM models that are trained for multi-task recommendation. We report the baselines for each of the three recommendation tasks in Appendix~\ref{appendix:baselines}. For fair comparison, we report the metrics of the T5-small variants of P5, VIP5, POD, and RDRec.

\subsection{Evaluation Metrics}
\label{subsec:evaluation_metrics}
For sequential and top-n recommendation, we evaluate models using Hit Rate (HR)@k and normalized discounted cumulative gain (NDCG)@k where $k \in \{1, 5, 10\}$. For explanation generation, we evaluate using the F1 scores of BLEU-4~\cite{papineniBLEUMethodAutomatic2002} and ROUGE-1, ROUGE-2, and ROUGE-L~\cite{linAutomaticEvaluationSummaries2003}. In all tables, the best performing model for each metric is in \textbf{bold}, the second best model is \underline{underlined}, and \textbf{*} indicates significance (p < 0.05) using one sample Student's t-test.

\subsection{Implementation Details}
\label{subsec:implementation_details}

To initialize user embeddings for \OURS{}, we train a probabilistic matrix factorization model~\cite{mnihProbabilisticMatrixFactorization2007} for top-n recommendation. We train for 100 epochs with a dimension size of 512, learning rate of 1e-3, and a lambda regularization of 1e-3.

After generating the embeddings, we train \OURS{} for 30 epochs. We utilize the AdamW optimizer~\cite{kingmaAdamMethodStochastic2015}, with a batch size of 64, learning rate of 5e-3, 4 attention heads, and the length of the task-specific prompt set to 3. We save the best performing model on the validation set from our hyperparameter search, and we report the mean score over 10 different seeds. 

 For Sports and Beauty, the optimal hyperparameters found were 20 similar users and the length of the collaborative user prompt $|\boldsymbol{p}_{mha}|=3$. For Toys, the optimal hyperparameters found were 10 similar users and $|\boldsymbol{p}_{mha}|=3$.  During inference, the number of beams is set to 20 and the inference batch size is set to 32 for all tasks. For the full implementation details, see Appendix~\ref{appendix:hyperparameter_details}.

\begin{table*}[!htbp]
\centering
\caption{Performance comparison on explanation generation, evaluated using BLEU (B) and ROUGE (RG).}
\label{tab:explanation_generation_comparison}
\resizebox{\textwidth}{!}{
\begin{tabular}{lcccc|cccc|cccc}
\hline
\multirow{2}{*}{Methods} & \multicolumn{4}{c}{Sports} & \multicolumn{4}{c}{Beauty} & \multicolumn{4}{c}{Toys} \\
\cline{2-5} \cline{6-9} \cline{10-13}
 & B-4 & RG-1 & RG-2 & RG-L & B-4 & RG-1 & RG-2 & RG-L & B-4 & RG-1 & RG-2 & RG-L \\
\hline
Att2Seq & 0.5305 & 12.2800 & 1.2107 & 9.1312 & 0.7889 & 12.6590 & 1.6820 & 9.7481 & 1.6238 & 13.2245 & 2.9942 & 10.7398 \\
NRT & 0.4793 & 11.0723 & 1.1304 & 7.6674 & 0.8295 & 12.7815 & 1.8543 & 9.9477 & 1.9084 & 13.5231 & 3.6708 & 11.1867 \\
PETER & 0.7112 & 12.8944 & 1.3283 & 9.8635 & 1.1541& 14.8497 & 2.1413 & 11.4143 & 1.9861 & 14.2716 & 3.6718 & 11.7010 \\
P5 & \underline{1.0447} & \textbf{14.9048} & \underline{2.1297} & \textbf{11.1778} & \underline{1.2237} & \underline{17.6938}  & \underline{2.2489} & \underline{12.8606} & 2.2892 & \textbf{15.4505} & \underline{3.6974} & \textbf{12.1718} \\
VIP5 & \textbf{1.0639} & \underline{14.8628} & 2.1012 & 11.1059 & \textbf{1.2850} & \textbf{17.7492} & \textbf{2.3482} & \textbf{12.9170} & \underline{2.3241} & \underline{15.3006} & 3.6590 & \underline{12.0421} \\
POD & 1.0013 & 14.0168 & \textbf{2.0436} & \underline{11.1236} & 1.0630 & 15.2517 & 1.5737 & 11.3283 & 2.3053 & 12.2889 & 
3.8512 & 10.3923 \\
\hline
PeaPOD &
1.0178 & 13.4364 & 1.8515 & 10.5017 & 
0.7933 & 12.2211 & 1.4744 & 9.1253 
& \textbf{2.5319} & 14.0476 & \textbf{4.0552} & 11.7318 \\

\hline
\end{tabular}
}
\end{table*}

\section{Results}
\label{sec:results}
\subsection{Recommendation Performance (RQ1)}

\textbf{Sequential Recommendation.}\quad In Table~\ref{tab:sequential_performance_comparison}, we compare our model with baseline models on the sequential recommendation task. Our results demonstrate that \OURS{} significantly outperforms all baselines on Sports and Toys, and performs comparably to IDGenRec on Beauty, though improvements are more modest compared to the other datasets. \OURS{}'s performance shows that user embeddings generated from one recommendation task can act as a strong signal for other recommender tasks. Although the user embeddings are generated from PMF, a top-n recommendation model, \OURS{} consistently outperforms the baseline models. This suggests that PeaPOD generalizes user preference patterns from collaborative filtering that transfer across recommendation paradigms.

\noindent\textbf{Top-n Recommendation.}\quad In Table~\ref{tab:topn_performance_comparison}, we report the performance of our method and the baselines on the top-n recommendation task. We find that \OURS{} outperforms existing baselines on 11 of the 12 metrics. In particular, PeaPOD shows significantly improved performance on Toys. Although \OURS{} was initialized with a variant of matrix factorization, \OURS{} showed significant gains over the matrix factorization baseline. Thus, our attention-based compositional strategy is able to further refine the collaborative information provided by the initial user embedding.

\noindent\textbf{Explanation Generation.}\quad
In Table~\ref{tab:explanation_generation_comparison}, we report the performance of all models on the explanation generation task. We observe that \OURS{} performs competitively on Toys and Sports, achieving the best performance in BLEU-4 and ROUGE-2 on Toys. However, \OURS{} performs worse on Beauty.
These mixed results reveal a crucial factor: the top-performing models, P5 and VIP5, incorporated \textit{additional item metadata} in their inputs that other models lacked. Our findings reveal that this item metadata, rather than collaborative user information, emerges as the critical factor for explanation generation quality. Specifically, user-based collaborative filtering provides insufficient signal for tasks requiring detailed item knowledge and may actually diminish performance on tasks that require specific information about items.


\subsection{Effects of User Prompt (RQ2)}

\begin{figure}[t]
  \centering 
  \includegraphics[width=\columnwidth]{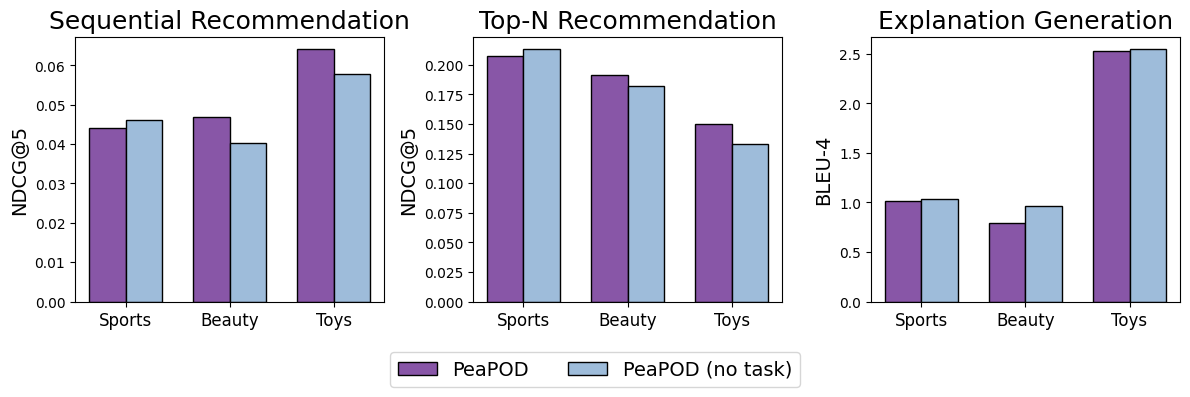}
  \caption{Comparison between PeaPOD and PeaPOD trained without task-specific prompts.}
  \label{fig:plot}
\end{figure}

\noindent\textbf{Training without task-specific prompt.}\quad To understand the effectiveness of collaborative user prompts versus globally shared prompts, we fine-tune \OURS{} with the same architecture and hyperparameters, but omit the task-specific prompts. We refer to the ablated model as \OURS{} (no task). We report our findings in Figure~\ref{fig:plot}.

We find PeaPOD performs better on Toys and Beauty, but slightly underperforms compared to PeaPOD (no task) on Sports. Crucially, both PeaPOD and PeaPOD (no task) outperform POD, meaning the personalized information distilled into the collaborative user prompt adds key information not captured by the task-specific prompt. Since both models are fine-tuned using the same base model and discrete prompt templates, we can attribute the gain in PeaPOD's performance to the effectiveness of the collaborative user prompt.

\begin{figure}[t]
\centering
  \includegraphics[width=\columnwidth]{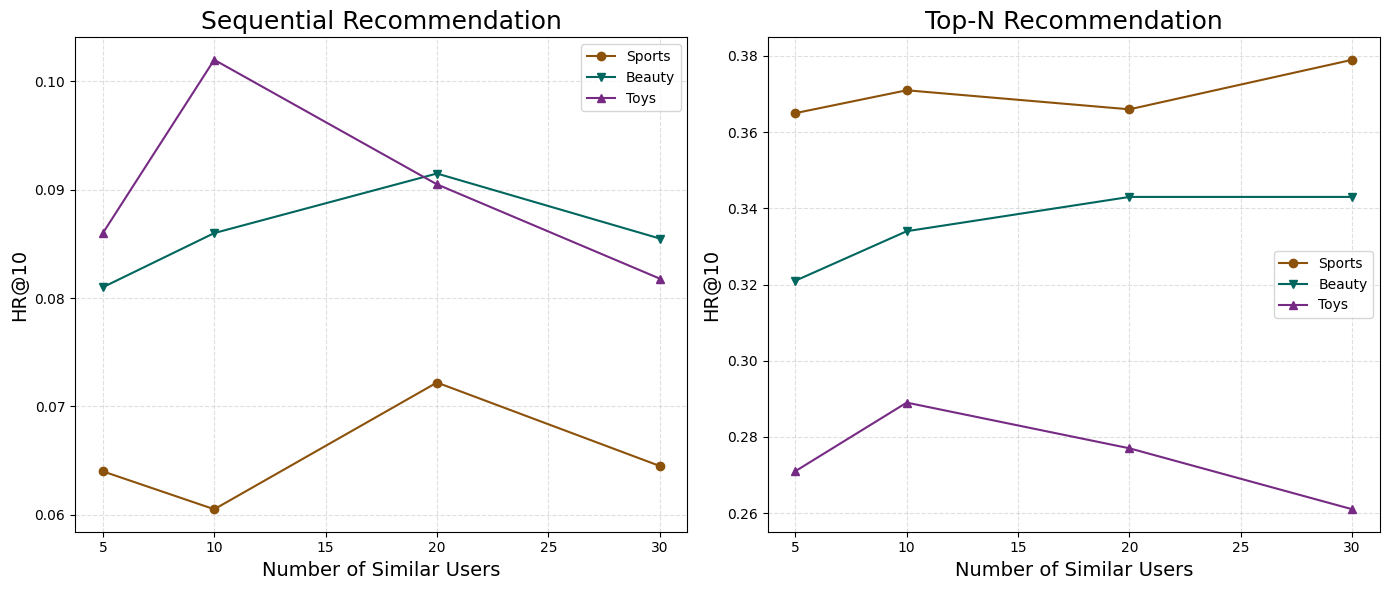}
  \caption {Performance comparison of sequential and top-n recommendation tasks with varying numbers of similar users.}
  \label{fig:neighbors}
\end{figure}

\noindent\textbf{Effects of number of neighbors.}\quad We investigated the effects of the number of similar users as part of our hyperparameter search in Figure~\ref{fig:neighbors}. We find that the number of similar users can affect the overall performance of the recommender and that this hyperparameter needs to be selected based on the dataset. Overall, we find that increasing the number of similar users benefits larger datasets, but the recommendation worsens if the number of neighbors is too high.

\noindent\textbf{Sensitivity to User Embeddings.}\quad
We train ablated versions of PeaPOD using different user embeddings to understand the effects of the initial user embeddings. In particular, we train PeaPOD with user embeddings generated from Bilateral Variational Autoencoder (BiVAE)~\cite{truongBilateralVariationalAutoencoder2021a} and LightGCN~\cite{lightgcn}.

\begin{figure}[t]
  \centering 
  \includegraphics[width=\columnwidth]{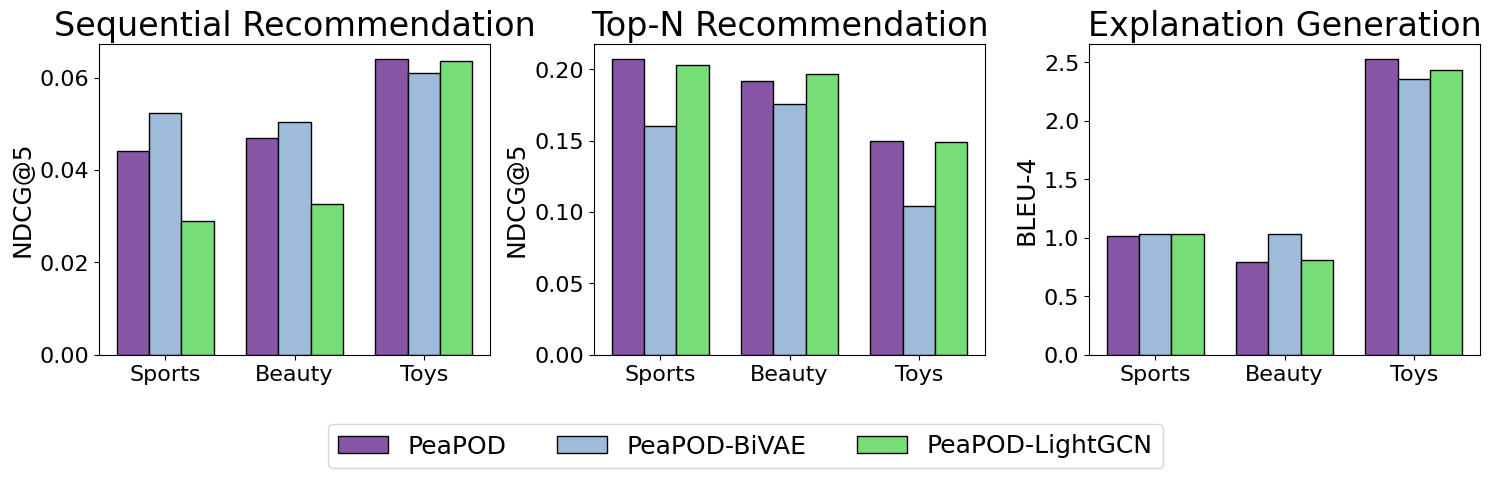}
  \caption {Comparison between PeaPOD trained with different user embeddings.}
  \label{fig:bivae_gcn}
\end{figure}

Figure~\ref{fig:bivae_gcn} shows that the method of initializing user embeddings affects how well the model performs on the different tasks. PeaPOD-BiVAE generally performed well on sequential recommendation, but PeaPOD and PeaPOD-LightGCN performed better in top-n recommendation.

Importantly, the user embedding \textit{does not} need to be generated from a traditional recommender system. We chose to initialize user embeddings using collaborative filtering because it provides key information about user-to-user interactions that the attention mechanism can further refine. However, alternative methods that capture personalized and collaborative preferences could potentially be used as input to \OURS{}.


\subsection{Effects of Multi-head Attention (RQ3)}
\label{subsec:multi_head_effects}

\begin{figure}[t]
\centering
  \includegraphics[width=\columnwidth]{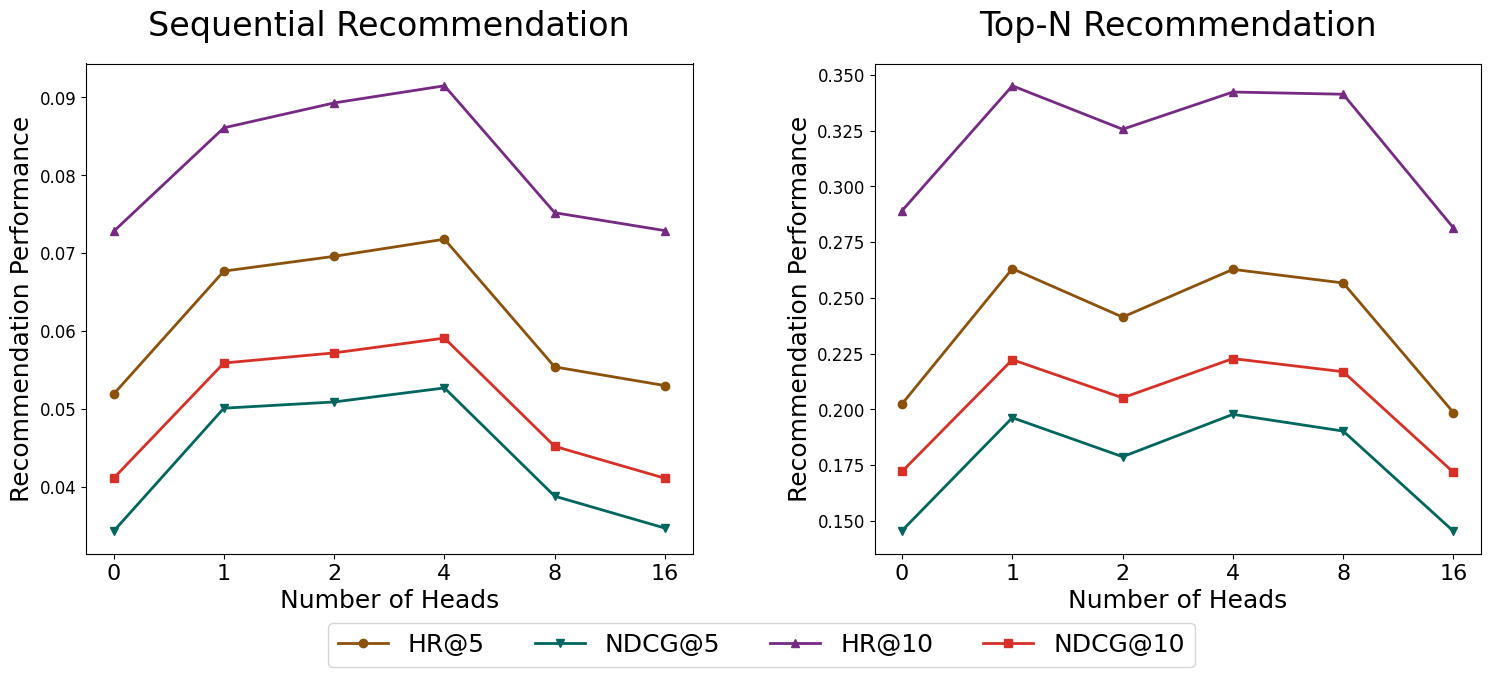}
  \caption {The effects of the number of attention heads on the Beauty dataset. Note that `0' represents MLP.}
  \label{fig:multihead}
\end{figure}

To understand the effects of multi-head attention versus alternative implementations, we performed a hyperparameter search on the Beauty dataset with varying numbers of heads as well as directly passing user embeddings through a multi-layer perceptron (MLP). We include MLP to observe whether attention provides superior performance compared to directly using a linear layer. We present the results in Figure~\ref{fig:multihead}.

We observe that multi-head attention with 4 heads, denoted as $MHA_4$, achieved the best overall performance, particularly in sequential recommendation. For top-n recommendation, the performance of $MHA_4$ and single-head attention were close to equal. We also observe that all attention-based models significantly outperformed the MLP implementation, showing the advantage of the compositional attention strategy. However, increasing the number of heads beyond 4 led to a decline in performance for all three recommendation tasks. This suggests that while adding more heads improves performance up to a certain point, exceeding this threshold causes the model to overfit.

\subsection{Effects of Model Size (RQ4)}
We investigate the effects of a larger base model size on the recommendation performance. In particular, we compare PeaPOD (small) and PeaPOD (base), which are models fine-tuned on T5-small (60.5M params) and T5-base (223M params), respectively. We report our findings in Table~\ref{tab:small_vs_base}.

We find that for both recommendation tasks, PeaPOD (base) performs similarly to PeaPOD (small) on Sports, but worse overall on Beauty and Toys. Our findings indicate that the smaller model performs better on smaller datasets and the difference between the two models decreases as the number of reviews scales. Thus, a bigger model does not always equate to better performance. Our findings are in alignment with \citet{gengRecommendationLanguageProcessing2022}, who also found that the P5-small variant outperformed the P5-base variant on the smaller datasets. The more competitive performance of PeaPOD (base) and P5 (base) on the Sports dataset suggests that a larger model may perform better as the number of user-item interactions increases.

\begin{table}[htbp] 
\centering 
\caption{Ablation between using T5-small and T5-base.}
\label{tab:small_vs_base} 

\resizebox{\columnwidth}{!}{
    \begin{tabular}{lrr|rr|rr}
    \hline
     & \multicolumn{2}{c}{Sports} & \multicolumn{2}{c}{Beauty} & \multicolumn{2}{c}{Toys} \\
    
     & HR@10 & N@10 & HR@10 & N@10 & HR@10 & N@10 \\
    \hline
    & \multicolumn{6}{@{}c}{Sequential recommendation} \\
    \hline
    
    PeaPOD (small) & \textbf{0.0665} & 0.0473 & \textbf{0.0843} & \textbf{0.0536} & \textbf{0.0917} & \textbf{0.0677} \\
    PeaPOD (base) & 0.0658 & \textbf{0.0488} & 0.0721 & 0.0488 & 0.0827 & 0.0610 \\
    
    \hline
    
    & \multicolumn{6}{@{}c}{Top-N recommendation} \\
    \hline
    PeaPOD (small) & \textbf{0.3645} & \textbf{0.2349} & \textbf{0.3417} & \textbf{0.2189} & \textbf{0.2793} & \textbf{0.1738} \\
    PeaPOD (base) & 0.3588 & 0.2300 & 0.3086 & 0.1914 & 0.2168 & 0.1291 \\
    \hline

    \end{tabular}
}
\end{table}

\section{Conclusion}
\label{sec:conclusion}
 In this paper, we propose PeaPOD, a novel generative recommender model that effectively distills individual and group user preferences into a collaborative user prompt. In particular, we introduce a multi-head attention strategy that uses the query-key-value mechanism to generate a personalized user embedding. Multi-head attention allows the model to capture multiple user-to-user relationships and enables the model to efficiently share collaborative knowledge between similar users. Our experiments confirm the effectiveness of our PeaPOD framework across three recommendation tasks. Future directions include incorporating additional metadata during training and developing an end-to-end architecture to unify the user embedding initialization and multi-head attention stages.

\section{Limitations and Ethical Concerns}
\label{sec:limitations}
One limitation of our approach is that we restrict our study to exclusively using user and item IDs. Thus, we do not incorporate any additional metadata such as user profiles or item descriptions. We find that this may hinder performance on explanation generation, which relies on having a strong understanding of both the user and item to provide a personalized response. We expect that providing additional metadata to the model can improve performance, particularly if it is given in natural language text, as the LLM is already pre-trained on such data. We plan to extend PeaPOD to utilize this information in future work. Additionally, PeaPOD relies on learning collaborative signals from the user embeddings to achieve good performance. Thus, similar to traditional collaborative filtering methods, PeaPOD may perform poorly in cold-start settings where there is little to no information about the user or item.

Furthermore, while our work focuses on benchmarking against other T5-based models, we acknowledge that newer decoder-only models have shown superior performance in NLP tasks. Due to computational limitations, we are unable to fine-tune larger SOTA models, but plan to do so in future work.

Finally, ~\citet{rajputRecommenderSystemsGenerative2023} raises potential concerns about data leakage in the original P5 data pipeline. In our work, all baseline methods in our comparison use the identical P5 preprocessing pipeline, ensuring a fair and consistent evaluation framework with existing works. The models we compare against are from recently published papers that report metrics on the same pipeline as ours. Our primary contribution is demonstrating that collaborative user prompts improve performance relative to existing methods under the same experimental conditions, but we acknowledge that \textit{all} baselines should be rerun to ensure no data leakage is taking place.


\section{Ethical Considerations}
The use of AI assistants (e.g. ChatGPT, Claude, etc.) was used to assist with coding and polishing the writing of this work.

\bibliography{main}

\appendix
\section{Baselines}
\label{appendix:baselines}
We report the following baselines for each of the three recommendation tasks:
\begin{itemize}
    \item For \textit{sequential recommendation}, we compare with the following baselines: GRU4Rec~\cite{jannachWhenRecurrentNeural2017}, \\BERT4Rec~\cite{sunBERT4RecSequentialRecommendation2019}, FDSA~\cite{zhangFeaturelevelDeeperSelfAttention2019}, SASRec~\cite{kang2018self}, P5~\cite{gengRecommendationLanguageProcessing2022}, VIP5~\cite{gengVIP5MultimodalFoundation2023}, POD~\cite{liPromptDistillationEfficient2023},
IDGenRec~\cite{tanIDGenRecLLMRecSysAlignment2024}, and RDRec~\cite{wang-etal-2024-rdrec}.
\vspace{-0.5em}
    \item For \textit{top-n recommendation}, we compare our method with: MF~\cite{korenMatrixFactorizationTechniques2009}, MLP~\cite{chengWideDeepLearning2016}, P5~\cite{gengRecommendationLanguageProcessing2022}, VIP5~\cite{gengVIP5MultimodalFoundation2023}, POD~\cite{liPromptDistillationEfficient2023},
and RDRec~\cite{wang-etal-2024-rdrec}.
\vspace{-0.5em}
    \item For \textit{explanation generation}, we use the following baselines: Att2Seq~\cite{dongLearningGenerateProduct2017}:, NRT~\cite{liGenerateNeuralTemplate2020a}, PETER~\cite{liPersonalizedTransformerExplainable2021a}, P5~\cite{gengRecommendationLanguageProcessing2022}, VIP5~\cite{gengVIP5MultimodalFoundation2023} and POD~\cite{liPromptDistillationEfficient2023}.
\end{itemize}

\section{Hyperparameter Details}
\label{appendix:hyperparameter_details}
 On average, each experiment took 6 hours to run. Since the main focus of our study is not on formulating discrete prompts, we reuse the same discrete prompts as POD~\cite{liPromptDistillationEfficient2023}. We adapt task-alternated training, as proposed by~\citet{liPromptDistillationEfficient2023}, to speed up training time by alternating each task per batch.

For the collaborative user prompt, we perform a hyperparameter search, testing prompt lengths of \{3, 5, 8, 10\}, number of heads of \{1, 2, 4, 8, 16\} and the number of similar users of \{5, 10, 20, 30\}. All models were trained on a Nvidia RTX 3090 GPU.

\section{Effects of User Prompts}

\subsection{Training without task-specific prompt}
We provide the full results of training without the task specific results in Tables~\ref{tab:ablation_sequential_performance_comparison},~\ref{tab:ablation_topn_performance_comparison}, and ~\ref{tab:ablation_explanation_generation_comparison}.

\begin{table*}[ht]
\centering
\caption{Performance comparison on sequential recommendation.}
\label{tab:ablation_sequential_performance_comparison}
\resizebox{\textwidth}{!}{
\begin{tabular}{lcccc|cccc|cccc}
\hline
\multirow{2}{*}{Methods} & \multicolumn{4}{c}{Sports} & \multicolumn{4}{c}{Beauty} & \multicolumn{4}{c}{Toys} \\
\cline{2-5} \cline{6-9} \cline{10-13}
 & HR@5 & N@5 & HR@10 & N@10 & HR@5 & N@5 & HR@10 & N@10 & HR@5 & N@5 & HR@10 & N@10 \\
\hline
PeaPOD
& 0.0554 & 0.0440 & 0.0665 & 0.0473 &
\textbf{0.0649} & \textbf{0.0469} & \textbf{0.0843} & \textbf{0.0536} & 
\textbf{0.0783} & \textbf{0.0641} & \textbf{0.0917} & \textbf{0.0677} \\
PeaPOD (no task) & 
\textbf{0.0582} & \textbf{0.0461} & \textbf{0.0711} & \textbf{0.0500} & 
0.0565 & 0.0402 & 0.0732 & 0.0456 & 
{0.0695} & 0.0579 & {0.0810} & 0.0612 \\
\hline
\end{tabular}
}
\end{table*}

\begin{table*}[t!]
\centering
\caption{Performance comparison on top-n recommendation.}
\label{tab:ablation_topn_performance_comparison}
\resizebox{\textwidth}{!}{
\begin{tabular}{lccccc|ccccc|ccccc}
\hline
\multirow{2}{*}{Methods} & \multicolumn{5}{c}{Sports} & \multicolumn{5}{c}{Beauty} & \multicolumn{5}{c}{Toys} \\
\cline{2-16}
 & HR@1 & HR@5 & N@5 & HR@10 & N@10 & HR@1 & HR@5 & N@5 & HR@10 & N@10 & HR@1 & HR@5 & N@5 & HR@10 & N@10 \\
\hline
PeaPOD & 
{0.1315} & {0.2784} & {0.2075} & {0.3645} & {0.2349} 
& \textbf{0.1217} & \textbf{0.2580} & \textbf{0.1915} & \textbf{0.3417} & \textbf{0.2189} 
& \textbf{0.0933} & \textbf{0.2033} & \textbf{0.1498} & \textbf{0.2793} & \textbf{0.1738} \\
PeaPOD (no task) & 
\textbf{0.1364} & \textbf{0.2849} & \textbf{0.2132} & \textbf{0.3692} & \textbf{0.2397} & 
{0.1130} & {0.2480} & {0.1820} & {0.3304} & {0.2079} & 
{0.0815} & {0.1821} & {0.1330} & {0.2575} & {0.1568} \\
\hline
\end{tabular}
}
\end{table*}

\begin{table*}[t!]
\centering
\caption{Performance comparison on explanation generation.}
\label{tab:ablation_explanation_generation_comparison}
\resizebox{\textwidth}{!}{
\begin{tabular}{lcccc|cccc|cccc}
\hline
\multirow{2}{*}{Methods} & \multicolumn{4}{c}{Sports} & \multicolumn{4}{c}{Beauty} & \multicolumn{4}{c}{Toys} \\
\cline{2-5} \cline{6-9} \cline{10-13}
 & BLEU-4 & ROUGE-1 & ROUGE-2 & ROUGE-L & BLEU-4 & ROUGE-1 & ROUGE-2 & ROUGE-L & BLEU-4 & ROUGE-1 & ROUGE-2 & ROUGE-L \\
\hline
PeaPOD &
1.0178 & 13.4364 & \textbf{1.8515} & {10.5017} & 
0.7933 & \textbf{12.2211} & 1.4744 & \textbf{9.1253} 
& {2.5319} & 14.0476 & {4.0552} & \textbf{11.7318} \\

PeaPod (no task) &
\textbf{1.0309} & \textbf{13.7973} & 1.8478 & \textbf{10.8804} &
\textbf{0.9598} &  11.3673 & \textbf{1.7317} & 8.8236 & 
\textbf{2.5487} & \textbf{13.8905} & \textbf{4.0814} & {11.6151} \\
\hline
\end{tabular}
}
\end{table*}

\subsection{Sensitivity to User Embeddings}
We provide the full results of training without the task specific results in Tables~\ref{tab:alt_sequential}, ~\ref{tab:alt_top_n}, and ~\ref{tab:alt_explanation}.

\section{Effects of Model Size}
We report the full results of training PeaPOD on T5-small vs T5-base in Tables~\ref{tab:base_sequential_performance_comparison}, ~\ref{tab:base_topn_performance_comparison}, and ~\ref{tab:base_explanation_generation_comparison}.

\begin{table*}[t]
\centering
\caption{Performance comparison on sequential recommendation, evaluated using Hit Rate (HR@k) and NDCG (N@k).}
\label{tab:alt_sequential}
\resizebox{\textwidth}{!}{
\begin{tabular}{lcccc|cccc|cccc}
\hline
\multirow{2}{*}{Methods} & \multicolumn{4}{c}{Sports} & \multicolumn{4}{c}{Beauty} & \multicolumn{4}{c}{Toys} \\
\cline{2-5} \cline{6-9} \cline{10-13}
 & HR@5 & N@5 & HR@10 & N@10 & HR@5 & N@5 & HR@10 & N@10 & HR@5 & N@5 & HR@10 & N@10 \\
\hline

PeaPOD
& 0.0554 & {0.0440} & {0.0665} & {0.0473} &
{0.0649} & {0.0469} & \textbf{0.0843} & {0.0536} & 
{0.0783} & \textbf{0.0641} & {0.0917} & {0.0677} \\
PeaPOD-BiVAE  & 
\textbf{0.0620} & \textbf{0.0524} & \textbf{0.0686} & \textbf{0.0542} &
\textbf{0.0677} & \textbf{0.0503} & {0.0835} & \textbf{0.0553} &  
{0.0719} & {0.0610} & {0.0801} & {0.0627} \\

PeaPOD-LightGCN  & 
{0.0456} & {0.0290} & {0.0643} & {0.0351} &
{0.0523} & {0.0326} & {0.0787} & {0.0411} &  
\textbf{0.0797} & {0.0635} & \textbf{0.0958} & \textbf{0.0678} \\
\hline
\end{tabular}
}
\end{table*}

\begin{table*}[t]
\centering
\caption{Performance comparison on top-n recommendation, evaluated using Hit Rate (HR@k) and NDCG (N@k).}
\label{tab:alt_top_n}
\resizebox{\textwidth}{!}{
\begin{tabular}{lccccc|ccccc|ccccc}
\hline
\multirow{2}{*}{Methods} & \multicolumn{5}{c}{Sports} & \multicolumn{5}{c}{Beauty} & \multicolumn{5}{c}{Toys} \\
\cline{2-16}
 & HR@1 & HR@5 & N@5 & HR@10 & N@10 & HR@1 & HR@5 & N@5 & HR@10 & N@10 & HR@1 & HR@5 & N@5 & HR@10 & N@10 \\
\hline
PeaPOD & 
\textbf{0.1315} & \textbf{0.2784} & \textbf{0.2075} & \textbf{0.3645} & \textbf{0.2349} 
& \textbf{0.1217} & {0.2580} & {0.1915} & {0.3417} & {0.2189} 
& \textbf{0.0933} & \textbf{0.2033} & \textbf{0.1498} & {0.2793} & {0.1738} \\
PeaPOD-BiVAE &
0.1020 & 0.2182 & 0.1602 & 0.2923 & 0.1830 & 
0.1133 & 0.2354 & 0.1758 & 0.3166 & 0.2012 & 
0.0590 & 0.1475 & 0.1041 & 0.2130 & 0.1246 \\
PeaPOD-LightGCN &
0.1312 & 0.2725 & 0.2033 & 0.3568 & 0.2298 & 
0.1211 & \textbf{0.2704} & \textbf{0.1970} & \textbf{0.3639} & \textbf{0.2264} & 
0.0909 & 0.2050 & 0.1491 & \textbf{0.2870} & \textbf{0.1751} \\
\hline
\end{tabular}
}
\end{table*}

\begin{table*}[!htbp]
\centering
\caption{Performance comparison on explanation generation, evaluated using BLEU (B) and ROUGE (RG).}
\label{tab:alt_explanation}
\resizebox{\textwidth}{!}{
\begin{tabular}{lcccc|cccc|cccc}
\hline
\multirow{2}{*}{Methods} & \multicolumn{4}{c}{Sports} & \multicolumn{4}{c}{Beauty} & \multicolumn{4}{c}{Toys} \\
\cline{2-5} \cline{6-9} \cline{10-13}
 & B-4 & RG-1 & RG-2 & RG-L & B-4 & RG-1 & RG-2 & RG-L & B-4 & RG-1 & RG-2 & RG-L \\
\hline
PeaPOD &
1.0178 & 13.4364 & 1.8515 & 10.5017 & 
0.7933 & 12.2211 & 1.4744 & 9.1253 
& \textbf{2.5319} & 14.0476 & \textbf{4.0552} & 11.7318 \\
PeaPOD-BiVAE & 
\textbf{1.0278} & \textbf{14.1174} & \textbf{2.0331} & \textbf{11.1607} & 
\textbf{1.0289} & \textbf{12.7236} & \textbf{1.8375} & \textbf{10.0963} &
{2.3544} & 12.7653 & {3.8807} & 10.7218 \\
PeaPOD-LightGCN &
\textbf{1.0278} & 13.8437 & 1.8915 & 10.8847 & 
0.8089 & 11.8196 & 1.4863 & 9.0550 
& {2.4356} & \textbf{14.5543} & {3.8975} & \textbf{12.0732} \\
\hline
\end{tabular}
}
\end{table*}

\begin{table*}[t]
\centering
\caption{Performance comparison on sequential recommendation between small and base models.}
\label{tab:base_sequential_performance_comparison}
\resizebox{\textwidth}{!}{
\begin{tabular}{lcccc|cccc|cccc}
\hline
\multirow{2}{*}{Methods} & \multicolumn{4}{c}{Sports} & \multicolumn{4}{c}{Beauty} & \multicolumn{4}{c}{Toys} \\
\cline{2-5} \cline{6-9} \cline{10-13}
 & HR@5 & N@5 & HR@10 & N@10 & HR@5 & N@5 & HR@10 & N@10 & HR@5 & N@5 & HR@10 & N@10 \\
\hline
PeaPOD (small)
& 0.0554 & 0.0440 & \textbf{0.0665} & 0.0473 &
\textbf{0.0649} & \textbf{0.0469} & \textbf{0.0843} & \textbf{0.0536} & 
\textbf{0.0783} & \textbf{0.0641} & \textbf{0.0917} & \textbf{0.0677} \\

PeaPOD (base) & 
\textbf{0.0562} & \textbf{0.0461} & {0.0658} & \textbf{0.0488} & 
0.0582 & 0.0443 & 0.0721 & 0.0488 &  
0.0731 & 0.0589 & 0.0827 & 0.0610 \\
\hline
\end{tabular}
}
\end{table*}

\begin{table*}[t]
\centering
\caption{Performance comparison on top-n recommendation between small and base models.}
\label{tab:base_topn_performance_comparison}
\resizebox{\textwidth}{!}{
\begin{tabular}{lccccc|ccccc|ccccc}
\hline
\multirow{2}{*}{Methods} & \multicolumn{5}{c}{Sports} & \multicolumn{5}{c}{Beauty} & \multicolumn{5}{c}{Toys} \\
\cline{2-16}
 & HR@1 & HR@5 & N@5 & HR@10 & N@10 & HR@1 & HR@5 & N@5 & HR@10 & N@10 & HR@1 & HR@5 & N@5 & HR@10 & N@10 \\
\hline
PeaPOD (small) & 
\textbf{0.1315} & \textbf{0.2784} & \textbf{0.2075} & \textbf{0.3645} & \textbf{0.2349} 
& \textbf{0.1217} & \textbf{0.2580} & \textbf{0.1915} & \textbf{0.3417} & \textbf{0.2189} 
& \textbf{0.0933} & \textbf{0.2033} & \textbf{0.1498} & \textbf{0.2793} & \textbf{0.1738} \\

PeaPOD (base) &
{0.1280} & 0.2698 & {0.2016} & 0.3588 & {0.2300} & 
0.1000 & 0.2268 & 0.1652 & 0.3086 & 0.1914 & 
0.0649 & 0.1494 & 0.1079 & 0.2168 & 0.1291 \\
\hline
\end{tabular}
}
\end{table*}

\begin{table*}[t]
\centering
\caption{Performance comparison on explanation generation between small and base models.}
\label{tab:base_explanation_generation_comparison}
\resizebox{\textwidth}{!}{
\begin{tabular}{lcccc|cccc|cccc}
\hline
\multirow{2}{*}{Methods} & \multicolumn{4}{c}{Sports} & \multicolumn{4}{c}{Beauty} & \multicolumn{4}{c}{Toys} \\
\cline{2-5} \cline{6-9} \cline{10-13}
 & BLEU-4 & ROUGE-1 & ROUGE-2 & ROUGE-L & BLEU-4 & ROUGE-1 & ROUGE-2 & ROUGE-L & BLEU-4 & ROUGE-1 & ROUGE-2 & ROUGE-L \\
\hline
\OURS{} (small) &
1.0178 & 13.4364 & 1.8515 & {10.5017} & 
0.7933 & 12.2211 & \textbf{1.4744} & {9.1253} 
& \textbf{2.5319} & 14.0476 & \textbf{4.0552} & {11.7318} \\

\OURS{} (base) & \textbf{1.0548} & \textbf{14.1369} & \textbf{2.0106} & \textbf{11.1413} & \textbf{0.8602} & \textbf{13.2299} & 1.3422 & \textbf{10.0028} & {2.5037} & \textbf{14.4358} & 3.9266 & \textbf{11.8846} \\
\hline
\end{tabular}
}
\end{table*}


\end{document}